\documentclass[sigconf]{acmart}

\usepackage{xcolor}
\AtBeginDocument{%
  }
\usepackage[normalem]{ulem}
\setcopyright{rightsretained}

\acmConference[]{Preprint}{revised May 21 2026}{under review}

\begin{document}

\title{Realizing Scaling Laws in Recommender Systems: A Foundation–Expert Paradigm for Hyperscale Model Deployment}

\makeatletter
\def\@affiliationfont{\large\normalsize}
\makeatother

\author{Dai Li*, Kevin Course*, Wei Li, Hongwei Li, Jie Hua, Yiqi Chen, Zhao Zhu, Rui Jian, Xuan Cao, Bi~Xue, Yu Shi, Jing Qian, Kai Ren, Matt Ma, Qunshu Zhang, Rui Li}
\thanks{\textsuperscript{*}Both authors contributed equally to this paper.}
\email{{daili1, kcourse, weilisjtu, lihw, mich94hj, yiqic, zhaozhu, rjian, xuancao, bixue, yushi2, jingqian, kren, zhenma, qunshuzhang, ruili}@meta.com}
\affiliation{%
  \institution{Meta Platforms, Inc.}
  \city{Menlo Park}
  \state{California}
  \country{USA}
}

\renewcommand{\shortauthors}{Li et al.}

\begin{abstract}
Scaling laws have been established for recommender systems, yet efficiently deploying foundation model (FM) across multiple recommendation surfaces remains a major unsolved challenge. Existing methods for transfer learning face fundamental limitations in this setting: knowledge distillation suffers from transfer fidelity degradation in the large-data regime, and static user or item embeddings lack the expressiveness to capture contextualized user-item interactions.

We propose the Foundation-Expert paradigm, where a central FM generates target-aware embeddings which are ingested by lightweight surface-specific expert models. Target-aware embeddings are representations that dynamically capture a user's interest in a specific item conditioned on their full interaction history. Unlike knowledge distillation, which transfers FM knowledge as soft labels, our approach provides these embeddings as input features to each expert model, enabling direct interaction with surface-specific representations.
This paradigm achieves transfer ratios of 0.64--1.0 from FM to experts, substantially exceeding existing methods.
Fully deployed at Meta serving tens of billions of daily requests since 2025, it delivers 0.050\% statistically significant online topline metric improvement and 0.359\% cumulative gains across multiple surfaces.

\end{abstract}

\ccsdesc[500]{Information systems~Recommender systems}

\keywords{foundation model, knowledge transfer, target-aware embedding, recommender system}

\received{21 May 2026}

\maketitle


\begin{figure}
    \centering
    \includegraphics[width=\linewidth]{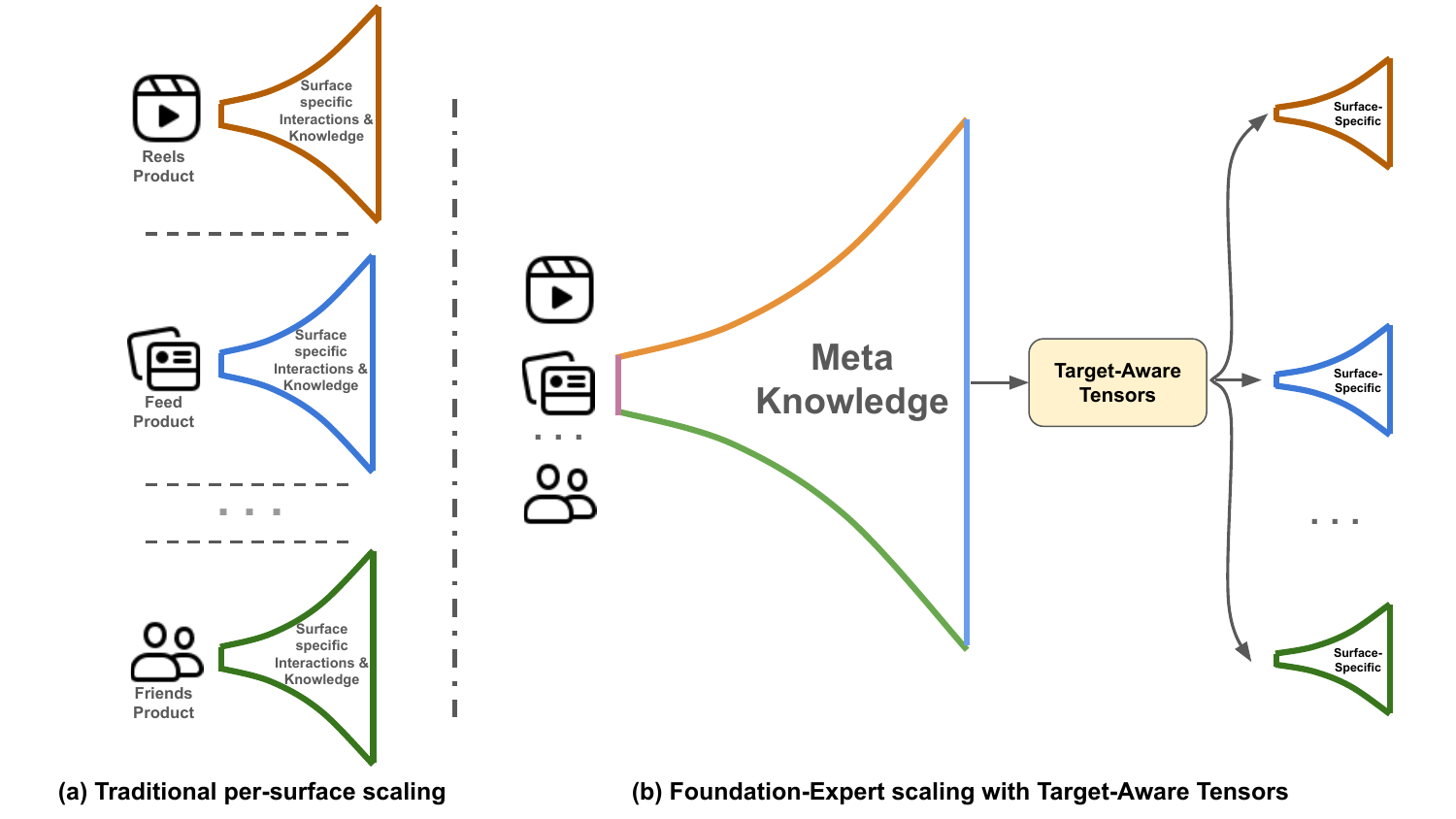}
    \caption{One-stage scaling (left) requires each surface to scale a monolithic model independently, duplicating effort. Our Foundation-Expert paradigm (right) trains one FM, then transfers knowledge via target-aware embeddings to lightweight, surface-specific experts, improving scaling efficiency and development velocity.}
    \label{fig:overview}
\end{figure}

\section{Introduction}

The systematic characterization of scaling laws has transformed deep learning practice~\cite{kaplan2020scaling}.
In recommender systems, recent work has established that model performance improves predictably with increased compute, data, and model scale~\cite{zhai_actions_2024,zhang2024wukong,han2025mtgr}, with architectural innovations further bending the scaling curve to achieve substantial compute efficiency gains~\cite{ding2026bending}.

Despite the profound potential offered by scaling recommender models,
their deployment in large-scale production environments presents a
significant challenge. First, training large recommendation models
often requires hundreds or even thousands of high-performance
GPUs, making efficient iteration challenging for researchers and
developers. Second, recommendation systems typically consist of multiple applications and surfaces, each requiring dedicated development and tuning. Attempting to independently scale, deploy, and maintain dedicated monolithic large models for every individual surface is computationally, financially, and operationally impractical.

A natural solution is to train/maintain a single foundation model (FM) and transfer its knowledge to surface-specific models/use cases. However, existing transfer learning paradigms face fundamental limitations in the streaming-data setting that characterizes industrial recommendation.
\begin{itemize}
    \item Supervised Fine-Tuning (SFT) is effective for foundation models (FMs) trained on static datasets~\cite{devlin_bert_2019,raffel_exploring_2020,hu_lora_2022}, it struggles in the context of recommendation systems. Industry scale recommendation system rely on  online training with one-epoch streaming data, where data distributions evolve constantly as new items surface and trends shift. In these non-stationary environments, SFT faces significant hurdles, including:catastrophic forgetting during the fine-tuning process~\cite{luo_empirical_2023}, performance degradation when adapting to shifting data distributions~\cite{kumar_fine-tuning_2022}, suboptimal update coordination between the foundation model and surface-specific models.
\item Knowledge distillation (KD)~\cite{hinton_distilling_2015,liang_external_2025, kd_google_transfer_ratio}—wherein a foundation model (teacher) generates soft labels to train surface-specific models (students)—is compatible with recommendation settings because both models can be continually updated. However, as demonstrated by our empirical evaluation (Section 4) and previous research studies~\cite{gou2021knowledge,busbridge_distillation_2025}, KD's knowledge translation ratio remains severely constrained, hovering around 10\%–25\%(refer Section~\ref{sec:results}). This limitation stems from several key factors: specialized loss functions are required to mitigate inherent teacher bias~\cite{gou2021knowledge}; student performance can be actively compromised by an overly capable teacher due to the capacity gap~\cite{busbridge_distillation_2025}. 

\item  In the recommendation domain, embedding-based methods~\cite{el2022twhin, zhang_scaling_2024, dv365, tencent, chen2025pinfm}, typically leverage a FM to independently learn static representations for a user based on the user's past behaviors (e.g., encoding long-term user preferences into a static embedding). Downstream, surface-specific models utilize these embeddings as input features. However, such static summaries fail to capture the fine-grained, dynamic, and contextualized interactions between a specific user and item, severely limiting their efficacy. Consequently, as demonstrated in our experiments, the knowledge translation ratio is a mere {25\%--30\%} (refer Section~\ref{sec:results}). 
\end{itemize}

To enable deploying a single Foundation Model (FM) across multiple recommendation surfaces, we propose a new \textit{Foundation-Expert paradigm}. Under this paradigm, a central FM generates target-aware embeddings that are then consumed as input features by lightweight, surface-specific expert models. Within each expert, these embeddings interact with surface-specific features and modules through jointly learned parameters, allowing every expert to optimize for the objectives and characteristics of its respective surface.

Technically, our paradigm design introduces two primary contributions:

\begin{enumerate}
    \item \textbf{Real-Time, Target-Aware Embeddings for High-Fidelity Knowledge Transfer:} 
    Unlike traditional static user embeddings that merely summarize general preferences, our target-aware embeddings capture the exact context of why a specific user interacts with a specific candidate item, conditioned on their full interaction history. The central FM generates these per-item/candidate embeddings in real time (on the order of minutes), enabling seamless, direct fusion with surface-specific features. This architecture achieves an unprecedented knowledge translation ratio of 0.64--1.0 from the FM to the experts, representing a $3\times$ improvement over standard knowledge distillation (KD) and static embedding baselines.

    \item \textbf{Cross-Surface Generalization via Raw User Interaction Signals:} 
    By training directly on raw user interaction history (UIH) signals and leveraging the FM's outputs as input features rather than target labels, the central model captures deeply generalizable knowledge. This facilitates a ``build once, deploy everywhere'' approach across multiple recommendation surfaces and tasks, including zero-shot scenarios for which the FM was never explicitly trained. Downstream expert models can then seamlessly ingest these rich features alongside their own customized surface data to optimize local performance.
\end{enumerate}

To support this paradigm, we develop HyperCast, a production-grade infrastructure system built for web-scale Foundation-Expert deployment. Our practical contributions include:

\begin{enumerate}
\setcounter{enumi}{2}
\item \textbf{System Design and Production Trade-offs:}
  We detail the end-to-end architecture of HyperCast, highlighting critical engineering practices across decoupled training, embedding freshness, and version management (Section~\ref{sec:hypercast-summary}). Particularly, decoupled training is necessary to maximize developer iteration speed,
  yet embedding freshness must be preserved despite real-world event-logging latencies to ensure high-fidelity knowledge transfer (Section~\ref{sec:results}).

\item \textbf{Large-Scale Empirical Validation:}
  We present comprehensive offline and online evaluations demonstrating the effectiveness of our framework across multiple surfaces. We share online A/B test results on a large-scale production recommendation platform at Meta with billions of daily active users. Our paradigm achieves a statistically significant 0.050\% topline metric lift (e.g., daily active users, app stay), driving a 0.359\% cumulative topline gain across all deployed surfaces. At our production scale, even a single-digit basis point (0.01\%) improvement represents a highly significant breakthrough.
\end{enumerate}

As a production proof of concept, we fully deployed this paradigm at Meta in 2025. A single central FM trained on 512 H100 GPUs concurrently serves knowledge to multiple surface-specific ranking models, with each expert requiring no more than $\sim$12\% of the FM's training accelerator footprint. Compared to scaling each surface's model independently, this paradigm reduces total training GPU resource requirements by $\sim3\times$ while drastically accelerating engineering velocity. To the best of our knowledge, this is the first real-world deployment of a real-time, target-aware Foundation-Expert paradigm at this scale in the recommendation industry, offering a proven, compute-efficient, and developer-friendly blueprint for realizing the promise of scaling laws across heterogeneous application surfaces.

\section{Related Works}\label{sec:related-works}

\paragraph{Long User History Modeling}
Over the past two years, much of the improvement in content recommendation quality has been driven by systems that learn from long user interaction histories---notably from Meta~\citep{zhai_actions_2024},
LinkedIn~\citep{hertel_efficient_2024},
ByteDance~\citep{chai_longer_2025},
Xiaohongshu~\citep{huang_towards_2025},
and Alibaba~\citep{wang_scaling_2025}.
Most recently, ULTRA-HSTU~\citep{ding2026bending} demonstrated over $5\times$ training and $21\times$ inference scaling efficiency gains through model-system co-design.
Our work is orthogonal to these architectural innovations: we focus on how to efficiently deploy the scaled model via a Foundation-Expert framework, and most such innovations can be incorporated into our FM design.
In this work, we leverage the HSTU architecture~\citep{zhai_actions_2024}.

\paragraph{Knowledge Transfer in Recommender Systems}
Knowledge distillation (KD) is a prevalent transfer paradigm, where a large teacher generates soft labels to train a smaller student~\cite{hinton_distilling_2015, liang_external_2025, kd_google_transfer_ratio}.
However, KD faces fundamental transfer fidelity challenges in the large-data regime~\cite{busbridge_distillation_2025} and requires specialized losses to mitigate teacher bias~\cite{gou2021knowledge}.
Alternatively, embedding-based methods learn general user~\cite{zhang_scaling_2024, dv365, tencent} or item~\cite{el2022twhin, pinsage, itemsage} representations for downstream consumption.
While computationally efficient, these static representations cannot capture the contextualized interaction between a specific user and a specific item, limiting transfer fidelity.
Recent work on target-aware modeling~\cite{zhai_actions_2024, chang2023twin, twinv2} has shown the importance of conditioning representations on both user history and the target item.
\citet{chen2025pinfm} developed an FM for learning from long user histories offline;
in contrast, our FM is trained in an online streaming setup with several-minute update frequency, generating target-aware embeddings for each candidate item.

\paragraph{Adapter and Expert Paradigms}
Adapter-based methods have demonstrated success across multiple domains, including computer vision~\cite{rebuffi_learning_2017} and NLP~\cite{pfeiffer_mad-x_2020}, with modular deep learning~\cite{pfeiffer_modular_2023} providing a theoretical grounding for composable, task-specific modules.
However, applying these paradigms to industrial recommender systems introduces unique challenges: models must learn from online streaming data under shifting distributions, adapt to diverse surfaces with heterogeneous objectives, and meet stringent latency constraints.
Our Foundation-Expert paradigm bridges this gap by adapting the adapter/expert concept to the streaming recommendation setting through target-aware embeddings.

\section{Methods} \label{sec:methods}
In this section, we introduce the design of our proposed Foundation-Expert paradigm, a two-stage architecture designed to overcome the inefficiencies in the traditional one-stage per-surface scaling of recommender systems.
A central, compute-intensive FM learns general knowledge from lifelong user histories, multi-modal content understanding, and cross-surface data.
The FM generates target-aware embeddings for each candidate item, which are consumed as input features by lightweight expert models.
This decoupling allows resource-intensive FM scaling and rapid expert iteration to occur in parallel.

\begin{figure*}
    \centering
    \includegraphics[height=2.8in]{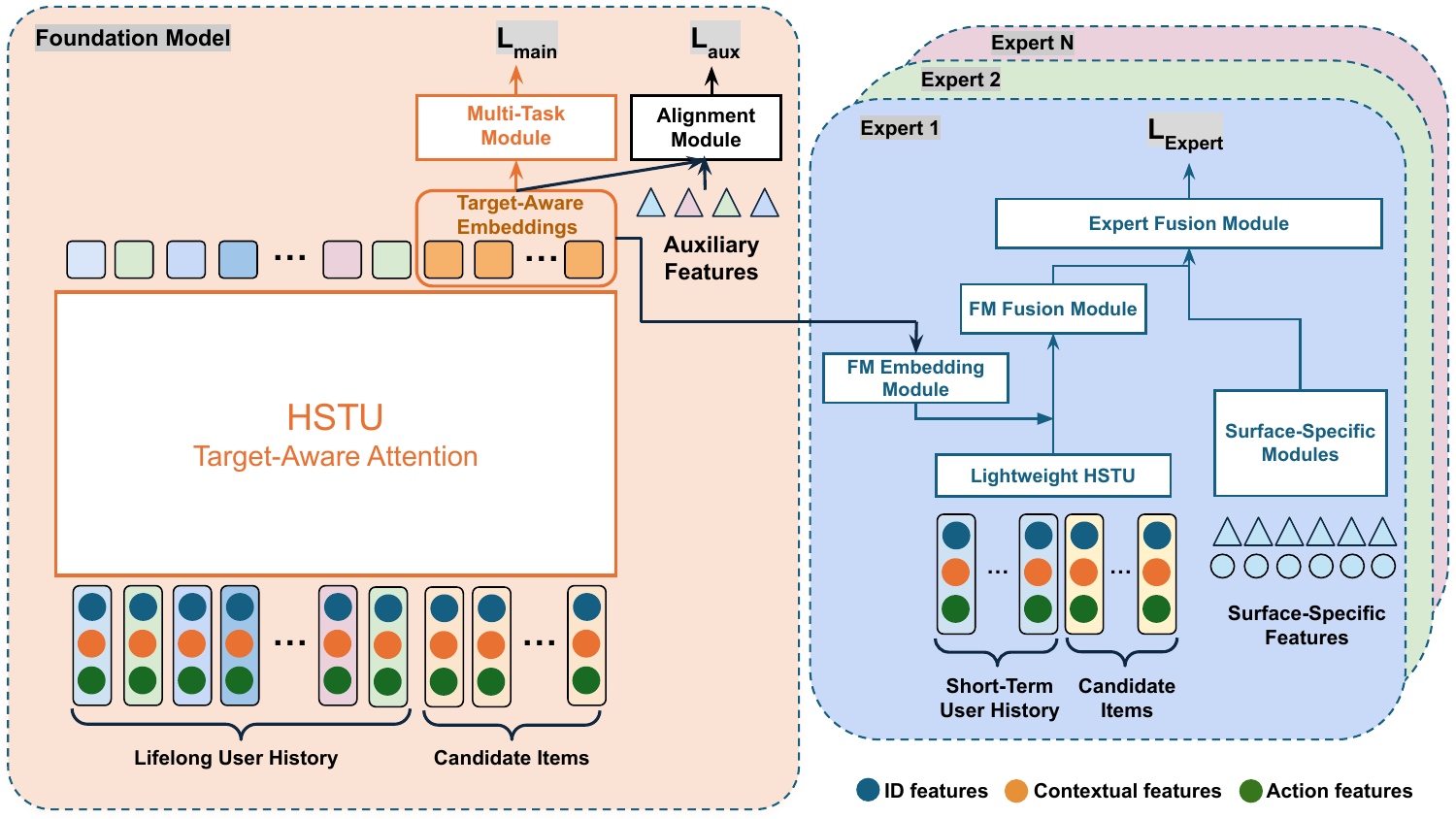}
    \caption{Overview of FM and Expert Model Architecture. The Foundation Model (FM) uses HSTU~\cite{zhai_actions_2024} to process lifelong, cross-surface user histories and candidate items, producing target-aware embeddings. These embeddings are then ingested by downstream expert models. Each expert uses its own lightweight HSTU to capture short-term, surface-specific signals. A FM Fusion Module combines the long-term knowledge from the FM embeddings with the expert's short-term representations. This fused embedding is then interacted with other surface-specific features to generate the final predictions.}
    \label{fig:arch}
\end{figure*}

\subsection{Foundation Model Design} 

\subsubsection{Input}
As depicted in Figure~\ref{fig:arch}, the FM is trained on a dataset comprising cross-surface, lifelong user histories and multi-modal content. Let $d$ denote the hidden dimension of the model. The input features are organized into two categories:

\textbf{Main Features} are used for target-aware sequential modeling to generate the FM embeddings.
These include the user's interaction history and information about the target items. Each item (historical or target) is represented by three groups of features: an item ID embedding $\mathbf{e}_p \in \mathbb{R}^{d_p}$, contextual feature embeddings $\mathbf{e}_c \in \mathbb{R}^{d_c}$ (including surface type, timestamp, and LLM-powered multi-modal representations), and an associated user action embedding $\mathbf{e}_a \in \mathbb{R}^{d}$.

\textbf{Auxiliary Features} consist of non-sequential data, such as common categorical, continuous and embedding features used in recommender systems.
These features, selected based on their importance in each surface, are used to aid the alignment of the FM embeddings during training for better generalizability on downstream experts.

\subsubsection{Target-aware Sequential Modeling}

The key property of target-aware embeddings is that they are conditioned on {both} the user's full interaction history {and} the specific candidate item.
This is achieved by appending candidate items to the user history sequence and processing them jointly through a sequential model, so that the resulting embedding for each candidate reflects the user's contextualized interest in that particular item.

To realize this, we leverage Hierarchical Sequential Transduction Units (HSTU)~\cite{zhai_actions_2024}, a transformer variant engineered for industrial-scale recommendation systems.
Building upon the original HSTU architecture, we introduce an architectural simplification depicted in Figure~\ref{fig:arch}: instead of interleaving item and action embeddings, we combine them via direct summation. 
This optimization effectively halves the input sequence length thereby yielding a 50\% reduction in complexity for the linear projection layers and a 25\% reduction for the attention operations.

Let $\mathcal{I}$ denote the finite catalog of all items across all recommendation surfaces.
The inputs consist of a chronologically ordered user history $(x_0, x_1, \ldots, x_{N-1})$ with $x_i \in \mathcal{I}$, and a set of $M$ candidate items $(y_0, y_1, \ldots, y_{M-1})$ with $y_j \in \mathcal{I}$, to be scored in a single request.
Let $f\colon \mathbb{R}^{d_p} \times \mathbb{R}^{d_c} \to \mathbb{R}^{d}$ be a multilayer perceptron that projects item and context embeddings into the hidden dimension $d$.
After initial preprocessing, we form a joint unified sequence of $d$-dimensional representations $(\mathbf{h}_{x_0}, \ldots, \mathbf{h}_{x_{N-1}},\, \mathbf{h}_{y_0}, \ldots, \mathbf{h}_{y_{M-1}})$, where each $\mathbf{h} \in \mathbb{R}^d$ is computed as:

\begin{equation}
    \mathbf{h}_{x_i} = f(\mathbf{e}_{p,i},\, \mathbf{e}_{c,i}) + \mathbf{e}_{a,i}
\label{eq:emb_1}
\end{equation}

\begin{equation}
    \mathbf{h}_{y_j} = f(\mathbf{e}_{p,j},\, \mathbf{e}_{c,j})
\label{eq:emb_2}
\end{equation}

Note that candidate items have no associated action embedding since they have not yet been interacted with.
With this unified sequence as input, sequential modeling in standard retrieval and ranking models can be formulated as shown in Figure~\ref{fig:arch}.

\subsubsection{Foundation Model Alignment}
Similar to many recommendation models, our FM is optimized using a multi-task multi-label (MTML) learning objective. The overall loss function $\mathcal{L}$ consists of two components,
\begin{equation}
    \mathcal{L} = \sum_{s=1}^{S} \omega_s \mathcal{L}^{\mathrm{main}}_{s} + \sum_{t=1}^{T} \omega_t \mathcal{L}^{\mathrm{aux}}_{t}
\label{eq:total_loss}
\end{equation}
where $S$ and $T$ denote the number of main and auxiliary tasks, respectively, $\mathcal{L}^{\mathrm{main}}_{s}$ and $\mathcal{L}^{\mathrm{aux}}_{t}$ denote the loss of main task $s$ and auxiliary task $t$, and $\omega_s, \omega_t \in \mathbb{R}_{>0}$ are the corresponding task weights.

\textbf{Main Loss ($\mathcal{L}^{\mathrm{main}}$)} This loss is derived from generalizable, cross-surface objectives such as likes, shares, and video completions. This supervision is applied directly to the HSTU module's output embeddings after a simple multi-task (MT) module, ensuring it can learn powerful and broadly applicable target-aware representations.

\textbf{Auxiliary Loss ($\mathcal{L}^{\mathrm{aux}}$)} This loss is designed for surface-specific alignment using crucial tasks from each domain. The target-aware embeddings are passed to a lightweight MLP (see ``Alignment Module" in Figure~\ref{fig:arch}) for interactions with auxiliary features.
To handle the heterogeneous nature of these tasks (e.g., engagement with video only happens on a product surface that presents videos), the loss for each auxiliary task is assumed to be zero when it is not available.

\subsection{Expert Design} 
In the Foundation-Expert paradigm, the traditional one-stage model for each production surface is replaced by a lightweight expert model. By offloading the compute-heavy task of general knowledge acquisition to the Foundation Model (FM), experts can be substantially smaller than their one-stage counterparts. This enables rapid iteration cycles focused exclusively on surface-specific optimizations.

 Our expert module consists of the following components: a FM Embedding Module, a FM Fusion Module, a lightweight sequence module (e.g., HSTU) dedicated to capturing short-term, real-time user interests and a expert fusion module.
 The data flow is as follows: 
 \begin{enumerate}
     \item The FM embedding module ingests the target aware embeddings and performs robustness enhancements (e.g., regularization, denoising), which are intended to allow the expert model to be more robust to FM embedding distribution shift. 
     \item The FM embeddings are fused with the output from the expert's lightweight sequence module through FM Fusion Module (e.g., a MLP). This enables the expert to fine-tune its target-aware representations on fresh surface-specific user history.
     \item Finally the fused embeddings are combined with representations from surface-specific modules (which are carefully tailored for specific recommendation surfaces) and fed into a standard MTML model to predict surface specific tasks.
 \end{enumerate}

The preceding design assumes that the FM's target-aware
embeddings are available to each expert at both training and
serving time.
We now describe the infrastructure that makes this possible.

\subsection{System Design: HyperCast}
\label{sec:hypercast-summary}
As discussed in Section~\ref{sec:methods}, the Foundation-Expert
paradigm decouples general knowledge acquisition from
surface-specific optimization.
Realizing this in a production environment serving tens of
billions of daily requests introduces infrastructure challenges
absent in monolithic models: training pipelines must be
synchronized without tight coupling, per-candidate embeddings
must be materialized and served at low latency, and both
components must be independently iterable.
To address these, we built HyperCast.
We share our
production design and the practical lessons learned from
deploying this paradigm at scale.

\subsubsection{Serving and Training Infrastructure}
At {serving time}, the FM generates target-aware
embeddings in real time for each candidate item in the user request.
These embeddings are consumed directly by the relevant
surface-specific expert within the same delivery flow,
resulting in end-to-end latency of $+1.2\%$ at $p50$,
$-1.6\%$ at $p95$, and $-2.2\%$ at $p99$ relative to the
one-stage baseline, with total CPU effectively unchanged
(within $0.02\%$ variance).

During {training}, FM and expert iterations are fully
decoupled.
The FM's target-aware embeddings, generated at inference time, are
logged as candidate-level features and made available in the
expert's training data.
Consequently, the FM and expert training jobs operate
independently, each consuming its own data and updating its own parameters without direct dependencies on the other's training
state.
We demonstrate the robustness of this decoupled design through successful online deployment and continued iterations in Section~\ref{sec:online-performance}, showing that expert models adapt robustly to FM embedding distribution drift.

Note that this design makes information leakage infeasible by construction.
Expert models can never be trained using embeddings generated for training
examples they have yet to see. 

\subsubsection{Freshness}\label{sec:freshness-design}
The FM is trained in online streaming
fashion, with model updates on the order of several minutes
and average data-to-trainer latency of approximately
30 minutes via a real-time event logging and joining pipeline.
This design choice is critical: as we validate empirically in
Table~\ref{tab:freshness},
relaxing the FM update frequency from minutes to hours 
drastically harms performance, confirming that the 
engineering investment in low-latency 
synchronization directly translates to model
quality.
Maintaining this level of freshness requires treating the
FM as a first-class production asset with dedicated monitoring,
on-call support, and automated rollback mechanisms.

\subsubsection{Version Management}
The decoupled architecture accelerates the development
lifecycle: experts can be iterated independently because FM
knowledge is materialized as input features, obviating the
need for joint training.
For rapid offline evaluation, HyperCast can load an FM
checkpoint into the expert's training flow to generate
embeddings inline, enabling quick experimentation without full
production deployment.
We also built a 
dedicated multi-version control framework that logs embeddings
from all active FM versions. Each expert is configured to
consume embeddings from a specific FM version, isolating model
lifecycles and enabling safe testing of various
Foundation-Expert combinations.

\subsubsection{Embedding Distribution Drift}
Version drift is a common challenge in industrial embedding-based systems. Our FM embeddings exhibit measurable distribution shift over time.
For example, we measured the drift in maximum mean discrepancy (MMD) with a Gaussian kernel for the embeddings generated by the FM at a reference time 
versus embeddings generated by the FM over the subsequent 5 hour period. 
After only 3 hours, the MMD between the embeddings and a random Gaussian reference is indistinguishable.
Despite the significant embedding drift, we find that the 
experts adapt robustly without instability in their output predictions.

The system design choices described above are what enable the
experimental validation in the proceeding section.

\section{Results} \label{sec:results}
In this section, we present a series of experiments to validate our proposed Foundation-Expert paradigm. We begin by demonstrating the effectiveness of the target-aware embeddings, the central component of our approach. Next, we show that performance improvements in the Foundation Model (FM) transfer effectively to expert models across multiple recommendation surfaces, and we analyze the generalization capabilities of the embeddings on tasks for which the FM was not explicitly trained. Finally, we present results from online A/B tests to validate the paradigm's feasibility and performance in a live production environment.

\subsection{Experiment Setup}

\paragraph{Data}
All experiments were conducted on industrial datasets.
Since this work required a tight coupling between infrastructure and modeling improvements to ensure the practical relevance and scalability, we did not apply our approach to public benchmarks.

\paragraph{Evaluation Metrics}
In the present work, we estimate offline performance using Normalized Entropy (NE), defined as the standard cross-entropy loss normalized by the entropy of the label distribution~\cite{he2014practical}. Since the denominator is a dataset-dependent constant, NE is directly proportional to log-loss and preserves the same ordering over models, while providing improved comparability across tasks with different positive rates. Given $n$ evaluation examples with binary labels $\ell_i \in \{0, 1\}$ and predicted probabilities $\hat{p}_i \in (0,1)$ for $i=1,\ldots,n$, the NE is defined as
\begin{equation}
    \mathrm{NE} = \frac{\frac{1}{n} \sum_{i=1}^{n}\bigl[\ell_i \log \hat{p}_i + (1- \ell_i) \log (1 - \hat{p}_i)\bigr]}
    {\bar{\ell}\, \log \bar{\ell} + (1-\bar{\ell}) \log (1-\bar{\ell})}
\end{equation}
where $\bar{\ell}=\frac{1}{n}\sum_{i=1}^{n} \ell_i$ is the base positive rate. We note that NE is Meta's primary offline metric for recommendation models across organic and ads surfaces, chosen for its strong correlation with online performance. In our products, 
 a relative improvement of $\approx 0.05\%$ is considered significant.
In Table~\ref{tab:ue_comp} we report both NE and AUC to demonstrate that they are directionally aligned. Subsequent experiments report NE exclusively, as the relationship between NE and online metrics is well-characterized and NE serves as a basis for all production launch decisions.

In the offline evaluation we assess model performance across several important tasks: 
(i)~``video complete" which indicates whether or not a user watches a video from start to finish;
(ii)~``like" and ``share" both of which are self-explanatory;
(iii) In addition to these broadly applicable tasks for both FM and most experts, the evaluation also incorporates surface-specific critical tasks, which follow the naming scheme of "Surface\_X\_Task\_i".

\paragraph{Foundation Model} The FMs evaluated in this study are trained in standard online streaming setup, utilizing data from four important recommendation surfaces. We evaluate two FM variants, designated HSTU-0.5B and HSTU-1B. It is important to note that the 0.5B and 1B model sizes here refer exclusively to the dense parameters; when including the sparse embedding tables the models operate on a trillion-parameter scale. We choose a target aware embedding dimension of 1024 per candidate item. Dense parameters are optimized with AdamW (lr=$4\times10^{-4}$, gradient clipping at 1.0). Training is conducted on 160 and 512 NVIDIA H100 GPUs for the HSTU-0.5B and HSTU-1B models, respectively. To enhance data freshness, per-surface downsampling is employed on the training data.

\paragraph{Experts} Each expert uses a lightweight HSTU module for surface-specific sequence processing, trained with no more than 12\% of the accelerators used by the HSTU-1B FM. The FM Fusion Module utilizes a simple MLP as a robust baseline. While more advanced fusion strategies may yield further improvements, an exploration of these is beyond the scope of this work. Similar to the FM, the experts also utilize data downsampling; however, the specific ratios for each expert are tailored to individual surface requirements and differ from those of the FM.

\subsection{Effectiveness of Target-Aware Embeddings}
Embedding-based and knowledge distillation (KD)-based methods are prevalent knowledge transfer paradigms in industrial recommender systems.
As discussed previously, both approaches have their limitations in the online streaming data settting. 

In this section we conduct an ablation study comparing our target-aware embeddings against the strongest internal user embeddings and real-time KD scores. 
For a fair comparison: (1)~the baseline user embeddings are produced by an user FM trained on the same user history time-range as our HSTU-1B FM, with 32× larger embedding dimensionality and multi-hour freshness, fused via target-aware attention (adding 5–7\% training overhead); (2) KD uses HSTU-1B FM as teacher to generate soft labels with the same freshness as our target-aware embeddings. The baseline excludes any FM-derived information.

As shown in Table~\ref{tab:ue_comp}, target-aware embeddings yield substantial NE improvements across all tasks, significantly outperforming both user embeddings and KD. The key distinction with KD is that our paradigm treats FM knowledge as input features rather than soft labels, enabling direct interaction with the expert's surface-specific representations—effectively adding compute and context in a way that soft-label mimicry cannot.

\begin{table}[h]
\centering
\begin{tabular}{lcc cc cc}
\toprule
\textbf{Model} & \multicolumn{2}{c}{\textbf{Like}} & \multicolumn{2}{c}{\textbf{Share}} & \multicolumn{2}{c}{\textbf{VC}} \\
\cmidrule(lr){2-3} \cmidrule(lr){4-5} \cmidrule(lr){6-7}
 & NE & AUC & NE & AUC & NE & AUC \\
\midrule
Baseline & 0 & 0 & 0 & 0 & 0 & 0 \\
UE & -0.64 & +0.12 & -1.15 & +0.32 & -0.78 & +0.41 \\
KD & -0.51 & +0.09 & -0.40 & +0.11 & -0.48 & +0.24 \\
\textbf{TAE (ours)} & -2.13 & +0.45 & -3.02 & +0.88 & -2.96 & +1.44 \\
\bottomrule
\end{tabular}
\caption{\% NE and AUC improvement relative to Baseline under our proposed target-aware embeddings (TAE). UE, KD, and VC denote user embedding, knowledge distillation, and video complete, respectively. A NE difference of $> 0.05\%$ is considered significant.}
\label{tab:ue_comp}
\end{table}

\paragraph{Embedding Freshness}
As discussed in Section~\ref{sec:freshness-design}, embedding freshness is a critical system design parameter.
Table~\ref{tab:freshness} quantifies the impact: relaxing the FM update frequency from several minutes to 1, 6, and 24 hours results in progressively larger NE regressions across all tasks, with video complete degrading by up to ${\sim}0.55\%$ at 24-hour staleness.
Notably, even with stale embeddings, the expert models retain reasonably robust performance in production. We attribute this robustness to their ability to synthesize general knowledge from the FM with their own surface-specific optimizations, granting them a degree of operational independence.

\begin{table}[h]
\centering
\begin{tabular}{lc}
\toprule
\textbf{FM Update Frequency} & \textbf{Video Complete NE Diff (\%)} \\
\midrule
Several minutes (baseline) & 0 \\
1 hour & +0.10 \\
6 hours & +0.27 \\
24 hours & +0.55 \\
\bottomrule
\end{tabular}
\caption{Impact of FM embedding staleness on video complete NE. Relaxing update frequency from minutes to hours results in progressive degradation, validating the freshness design in Section~\ref{sec:freshness-design}.}
\label{tab:freshness}
\end{table}

\subsection{Foundation-to-Expert Transfer Efficiency} \label{sec:transfer-ratio-experiment}
A key advantage of our paradigm is its transfer efficiency: the FM can be improved centrally, with gains propagating at high ratio to numerous downstream experts simultaneously.

To quantify this, we conducted an experiment training two architecturally identical expert models, differing only in their input FM embeddings: one uses the HSTU-0.5B FM and the other the HSTU-1B FM. Both experts are warmed up from a model trained on HSTU-0.5B embeddings for over one month. We define the Transfer Ratio (TR) between a weaker FM$_1$ and a stronger FM$_2$ (i.e., $\mathrm{NE}(\text{FM}_1) > \mathrm{NE}(\text{FM}_2)$) for a given expert as
\begin{equation}
    \mathrm{TR} = \frac{\mathrm{NE}(\text{Expert}_{\text{FM}_1}) - \mathrm{NE}(\text{Expert}_{\text{FM}_2})}{\mathrm{NE}(\text{FM}_1) - \mathrm{NE}(\text{FM}_2)}
\end{equation}
so that both numerator and denominator are positive when the stronger FM improves expert performance. $\mathrm{TR}$ measures the proportional expert improvement relative to the underlying FM improvement; higher values indicate more efficient transfer.

As shown in Table~\ref{tab:transfer-ratio-surf}, our approach achieves transfer ratios in the range of $[0.64, 1.0]$, substantially exceeding user/item embedding-based and KD-based solutions internally in Meta's recommender stacks. 
Note that $\mathrm{TR} \geq 1$ can occur when improved FM embeddings interact synergistically with the expert's surface-specific features, amplifying gains beyond the raw FM improvement. These results confirm that FM scaling gains propagate efficiently to downstream experts, eliminating the need to scale each surface model independently.

\begin{table}[h]
\centering
\begin{tabular}{l l l c  }
\toprule
\textbf{Surface} & \textbf{Task Type} & \textbf{Task Name} &\textbf{Transfer Ratio} \\
\midrule
\textbf{Surface A} & Main & Like & 0.7397 \\
 & Main & Share & 1.0000 \\
 & Main & Video Complete & 0.9060 \\
 \midrule
\textbf{Surface B} & Main & Like & 0.7228 \\
 & Main  & Share & 0.8000 \\
 & Main & Video Complete  & 0.6765 \\
 & Aux & Surface\_B\_Task\_1  & 0.6437 \\
 & Aux & Surface\_B\_Task\_2  &  0.8932 \\
 & Aux & Surface\_B\_Task\_3  &  0.7361 \\
 \midrule
\textbf{Surface C} & Main & Like & 0.7792 \\
 & Main & Share & 0.9020 \\
 & Main & Video Complete & 0.7236 \\
 & Aux & Surface\_C\_Task\_1 & 0.9231 \\
 & Aux & Surface\_C\_Task\_2 & 0.9302 \\
 & Aux & Surface\_C\_Task\_3 & 0.8250 \\
\bottomrule
\end{tabular}
\caption{Foundation-to-Expert Transfer Efficiency across Surfaces. For "Task Type", "Main" means that task is main task for both FMs and Experts. "Aux" means that task is auxiliary task for FMs while main task for Experts.  
}
\label{tab:transfer-ratio-surf}
\end{table}

\subsection{Generalization to Unseen Tasks} \label{sec:generalizability}
While the previous experiments in Section~\ref{sec:transfer-ratio-experiment} demonstrated FM's strong generalizability across surfaces, the FM has been exposed to all the surface-specific tasks as either main or auxiliary objectives.  In this section, we investigate a more challenging scenario: the FM's ability to generalize to expert tasks on which it has no direct training supervision. 

For this experiment, we established a baseline using the production model of "Surface D" without our FM embeddings. The expert model is architecturally identical to the baseline but incorporates embeddings from the HSTU-0.5B FM. Notably, the FM was trained using around 20\% of the "Surface D" data, in contrast to the baseline which was trained on the full 100\%. And this FM was aligned using only one of the primary tasks from Surface D as an auxiliary objective. We then measured expert performance relative to the baseline on the four other tasks from Surface D, which were intentionally withheld from the FM's training.

As shown in Table~\ref{tab:general}, the expert model with FM embeddings achieved statistically significant gains on all four "unseen" tasks over the baseline. This result underscores the FM's powerful generalization ability, proving it can learn and transfer knowledge that is broadly useful, even for tasks beyond its explicit optimization objectives. This capability is a cornerstone of our "build once, use everywhere" vision, enabling a single FM to benefit an entire ecosystem of diverse and evolving tasks.

\begin{table}[h!]
    \centering
    \begin{tabular}{lc}
        \toprule
        \textbf{Task Name} & \textbf{NE Diff (\%) vs.\ Baseline} \\
        \midrule
        Surface\_D\_Task\_1 & -0.60 \\
        Surface\_D\_Task\_2 & -0.53 \\
        Surface\_D\_Task\_3 & -0.40 \\
        Surface\_D\_Task\_4 & -0.51 \\
        \bottomrule
    \end{tabular}
    \caption{Expert performance improvements on Surface D tasks that were not seen by the FM during its training. The baseline is the production model.}
    \label{tab:general}
\end{table}

\subsection{Online Performance}\label{sec:online-performance}
We validated our paradigm through extensive online A/B tests on a large-scale recommendation platform at Meta, serving tens of billions of daily user requests. We report results in two settings. 
In the ``Benchmark" setting we directly compare the Foundation-Expert paradigm against a one-stage baseline where we serve the foundation model prediction directly on a single surface.
In the ``Enablement" setting, we report cumulative gains since the initial launch of the Foundation-Expert paradigm into production. 

We track three categories of online metrics: consumption (e.g., watch time), engagement (e.g., likes, comments, shares), and topline (e.g., app stay, daily active users). As summarized in Table~\ref{tab:online}, our paradigm achieves statistically significant improvements across all categories, including a notable engagement shift towards fresher content. 
We attribute these gains to the explicit separation of concerns. The FM captures general, long-term knowledge while the expert specializes in surface-specific optimization and real-time user intent. In our system, topline improvements of 0.05\% are highly significant. Since its launch in 2025, our paradigm has delivered 0.359\% cumulative topline improvement across multiple surfaces in this platform through efficient scaling and knowledge transfer.

\begin{table}[h!]
    \centering
    \begin{tabular}{llr}
        \toprule
        \textbf{Type} & \textbf{Online Metric} & \textbf{\% Improvement} \\
        \midrule
        \textbf{Benchmark} & Consumption & $1.079$ \\
        & Engagement & $1.63$ \\
        & Freshness & $2.749$ \\
        & Topline & $0.050$ \\
        \midrule
        \textbf{Enablement} & Consumption & $6.512$ \\
        & Engagement & $11.24$ \\
        & Freshness & $12.65$ \\
        & Topline & $0.359$ \\
        \bottomrule
    \end{tabular}
    \caption{Online A/B test results. \emph{Benchmark}: direct comparison of the two-stage Foundation-Expert paradigm against the one-stage baseline. \emph{Enablement}: cumulative gains since launch through successive FM iterations across multiple surfaces.}
    \label{tab:online}
\end{table}

\section{Conclusion}\label{sec:conclusion}
We introduced the Foundation-Expert paradigm, which decouples a central, compute-heavy FM from lightweight, surface-specific experts and transfers knowledge via \emph{realtime} \emph{target-aware embeddings}. Powered by the HyperCast infrastructure, this paradigm achieves transfer ratios of 0.64--1.0 and delivers statistically significant online improvements across multiple surfaces at Meta.

Key areas for future work include (i)~exploring richer fusion architectures and multi-modal expert designs and (ii)~extending the paradigm to additional recommendation surfaces and non-recommendation tasks.
Currently fully deployed across multiple core recommendation surfaces at Meta, this work provides a proven blueprint for realizing the benefits of scaling laws in complex, real-time recommendation environments.


\begin{acks}
This work represents the joint efforts of many of engineers, researchers, data scientists, and leaders and would not be possible without the following individuals (listed alphabetically): Banit Agrawal, Bin Kuang, Bugra Akyildiz, Chao Deng, Charlie Li, Chelsea Pan, Chenran Li, Chloe Liu, Chufeng Hu, Chunxing Yin, Colin Peppler, Cong Shen, Daisy Shi He, Franco Mo, Han Li, Hao Lin, Hao Wan, Hong Yan, Hongyi Jia, Hongzheng Shi, Huihui Cheng, Ilina Mitra, Jack Chai, James Zuo, Jeet Kanjani, Jiahao Luo, Jiaqi Zhai, Jing Ma, Jing Shan, Ke Gong, Lars Backstrom, Linjian Ma, Lu Fang, Lu Zhang, Marcus Gao, Meihong Wang, Michael Chen, Michael He, Mike Ching, Min Ni, Nikki Zhang, Ning Jiang, Pai-Wei Lai, Qianqian Zhong, Rajasi Saha, Ram Ramanathan, Rex Cheung, Rui Jian, Rui Zhang, Runming Lu, Shikha Kapoor, Shilin Ding, Shiyan Deng, Shouwei Chen, Siqiao Chen, Sophia (Xueyao) Liang, Wen-Yun Yang, Xiaoxing Zhu, Xinyao Hu, Xinye Zheng, Xudong Ma, Yanhong Wu, Yifan Shao, Yisong Song, Yuting Zhang, Zhe (Joe) Wang, Zhuoran Zhao, Zimeng Yang.
\end{acks}

\newpage
\bibliographystyle{ACM-Reference-Format}
\bibliography{paper}

\end{document}